# Combining Topic Modeling with Grounded Theory: Case Studies of Project Collaboration

*Eyyub Can Odacioglu (eyyub.odacioglu@manchester.ac.uk)*
*The University of Manchester*

*Lihong Zhang (lihong.zhang@manchester.ac.uk)*
*The University of Manchester*

*Richard Allmendinger (richard.allmendiger@manchester.ac.uk)*
*The University of Manchester*

## Abstract
This paper proposes an Artificial Intelligence (AI) Grounded Theory for management studies. We argue that this novel and rigorous approach that embeds topic modelling will lead to the latent knowledge to be found. We illustrate this abductive method using 51 case studies of collaborative innovation published by Project Management Institute (PMI). Initial results are presented and discussed that include 40 topics, 6 categories, 4 of which are core categories, and two new theories of project collaboration.

## Introduction

Theory building and new knowledge generation have always been attractive and challenging to management studies. Qualitative researchers, in particular, preferably use Grounded Theory (GT) and Case Study (CS) to complete this exploring process. Positivist methods toward GT and CS have gained acceptance (Creswell and Creswell, 2018; Morse et al., 2021). Subjectivist researchers, on the other hand, are struggling to progress due to the quality constraints and academic continuity - e.g. few scholars following previous studies in this risky business (Creswell and Creswell, 2018).

Köhler et al. (2022) questioned the fact that objectivist scholars over-rely on the prescribed procedures to construct new knowledge in GT (Gioia et al., 2013) and CS (Yin, 2018). These procedure-based linear processes are characterised as systemised, mechanical, simplified, pattern recognition and repeatable - so-called "templates". With a template, scholars analyse data by seeking patterns and building categories according to identified patterns, which leads to predefined generalisation. Templates are widely accepted and used because it's easy to follow the provided prescripts, and repeatability makes them close to positivism (Mees-Buss et al., 2022). Researchers are thus advised to adopt a well–known template and strictly follow it in their study.

These above linear processes provide validity, reliability (Creswell and Creswell, 2018), and procedural rigour to research (Harley and Cornelissen, 2022). However, they cause losing some vital dimensions of interpretivism (Köhler et al., 2022), such as innovation in the research design (Cilesiz and Greckhamer, 2022; Lê and Schmid, 2022), the interpretive rigour (Locke et al., 2022), and reducing research trustworthiness and transparency (Lerman et al., 2022; Pratt et al., 2022). Morse et al. (2021) criticised the template applied studies for concluding research by identifying patterns in the data without reaching analytical abstraction.

Current template-based methodologies, especially GT, have two drawbacks in the big data era. Firstly, these methodologies follow a linear process, and they lead studies to predefined generalisation overlooking the balanced approach of subjectivism (Locke et al., 2022). Secondly, they are not suitable for coding a vast amount of data. In classical qualitative research, particularly GT, a human has to analyse data and needs to master the data and context (Charmaz and Belgrave, 2019). Constructivist Grounded Theory (CGT) of Charmaz (2006) is one of the solutions for the first problem. In CGT, she emphasised that a researcher cannot reach generality with a prescribed goal for it. Besides, Morse et al. (2021) asserted that generality emerges from the analytic process, which involves interpreting, discovering, or constructing by applying analytic tools such as constant comparison and memo writing. A new computer-aided text analysis technique, Topic Modeling (TM), is a solution for coding big textual data in a shorter time (Baumer et al., 2017). Although TM has received increasing attention from scholars because of its novelty (Dimaggio, 2015; Lee et al., 2020), it only generates codes, and theory development still requires human interpretation in TM (Schmiedel et al., 2019). For this reason, researchers use TM with qualitative analysis methods such as discourse analysis (Aranda et al., 2021) and content analysis (Chuang et al., 2014). After TM generates topics (or codes), the theory is developed with pattern recognition (Bansal et al., 2020). Again, these studies can be considered TM added templates. Hence, the literature lacks a method that aids a novel theorising with a robust process that can utilize an algorithm to analyse a vast amount of textual data to overcome the shortcomings of existing template-based methodologies.

Therefore, we propose to combine the CGT of Charmaz (2006) with a TM algorithm of Blei et al. (2003), which provides both interpretive rigour and procedural rigour. We argue that this methodology is best for analysing and theorising dynamic phenomena, such as innovation, where several dynamic processes coincide, and a variety of analytical dimensions should be considered simultaneously (Crossan and Apaydin, 2010), and collaboration, where several stakeholders iteratively interact in an ecosystem of innovation-based projects (Odacioglu et al., 2021). (In this above situation, collaboration and innovation are interlocked in a complex development project.) It consists of four well-defined steps that firstly provide procedural rigour and fulfil the validity requirement. Secondly, these well-developed steps support the analytical abstraction of CGT, which provides hermeneutic rigour. Thirdly it enables us to explore and translate case studies published from a professional big data-based website - The Project Management Institute (PMI) - and finally, to answer the following questions:

RQ1: How can a Grounded Theory approach combine a computerised method of textual data analysis that enables human interpretation?

RQ2: What hidden topics and possible theories on the project collaboration (and innovation) have been discussed together in a professional management community?

The following section of the Literature Review briefly discusses the existing GT, CS and TM studies. Then the Methodology Section outlines data collection and analysis. The penultimate section discusses initial findings that lead to the final Conclusion.

**Literature Review**
Applying a novel technique is as important as exploring new knowledge for management researchers to make an impact in the field. For this reason, their adaptation rate to new techniques could be higher than any other scholars. Big data approaches were not an exception, and scholars rapidly adopted them. Even these approaches bring challenges with opportunities for traditional GT and CS searchers (Zhang et al., 2016). George et al. (2016) asserted that while big data requires a new skill set and effort, the scope and granularity of data provide possibilities of providing better answers to further questions to develop new theories attracted the management researcher. When it comes to qualitative research, because of its inductive nature, TM is the most preferred textual big data approach (Ignatow and Mihalcea, 2017).

However, TM is not a standalone method to apply and find new knowledge. When it is used alone, it can only help to make future predictions by identifying trends and patterns in the literature or in patent documents (e.g. Choi and Song (2018); Han et al. (2021); Moro et al. (2020)). It requires human interpretation for sense-making. Although we do not know what will happen in the future if AI will replace human interpretation or not, today, these kinds of tools are suggested as supportive tools, and scholars assert that they should not replace human interpretation (Odacioglu et al., 2021). For example, Schmiedel et al. (2019) emphasised TM as a complementary methodological approach for management research to help a researcher to analyse a vast amount of textual data for a management phenomenon. Scholars combined TM with qualitative text analysis methods, such as content analysis (e.g. Chuang et al. (2014), Piepenbrink and Gaur (2017)), discourse analysis (e.g. Aranda et al. (2021), Jacobs and Tschötschel (2019)), thematic analysis (e.g. Valdez et al. (2018)), and narrative analysis (e.g. Isoaho et al. (2021)). The study of Hannigan et al. (2019) showed different novel knowledge identification applications by using the TM, which highlighted TM as a text-based big data method for theory development and promising application for future management research. Hence, all of these studies proved that TM can facilitate contributing knowledge for management research.

Moreover, these contributions made a researcher search for possibilities to use TM with GT. For example, Nelson (2020) defined a linear procedural computer-based GT for a sociology study and used TM and other text analysis approaches to build their theory. In summary, it is a template like linear methodology, and she searched for patterns and categorised topics, then confirmed these patterns by utilising other computer-based pattern recognising techniques, such as word hierarchies, to identify linguistic relations. Baumer et al. (2017) compared and contrasted the GT with TM for management research. They simultaneously applied both GT and TM with content analysis to the same data set to highlight the similarities in data collection and analysis steps, procedures and outcomes.

Recently, Croidieu and Kim (2018) would be the only example which could integrate them. However, they stated that they followed the template of Strauss and Corbin (1997). Baumer et al. (2017) pointed out the need for the use of GT with TM together in a novel way. The limited studies and calls revealed that the literature lacks the template-free methodologies that combine them to highlight the interpretivist rigour. This is meaningful because Grounded Theorists are also looking for new methodologies that combine the theory-building power of GT with any big data approaches (Walsh et al., 2015).

Baumer et al. (2017) proved the technical similarity between TM and GT. However, the technical fit is not enough to merge the two methodologies. Besides, Jacobs and Tschötschel (2019) stress the epistemological, ontological and philosophical fit of combining two methodologies. We believe there are similarities between CGT and TM. For example, the significance of data, pragmatic roots, methodological steps to explore the (latent) meaning in data, and the requirement of a high level of human interpretation are all features that are shared by both TM and GT. Both methods start with inductive coding with exploratory purposes, and the human interpretation takes the study to abductive reasoning building (see Charmaz (2006) and Hannigan et al. (2019)). The technical, epistemological and ontological fit makes it possible to merge TM and GT. The following section will answer our first research question, and the Findings and Results Section will answer our second question.

**Methodology**

The previous sections discussed problems of current positivist oriented qualitative research and existing TM applications. Additionally, the literature emphasised the need for combining TM with GT in a novel way. After discussing the data and their means of collection, this section illustrates the steps of the proposed methodology. The methodology is designed to overcome the shortcomings of existing linear methodologies, such as reaching predefined theory and the capability to analyse big textual data. It is an analytic process to identify the latent meanings in the text and hidden knowledge in the data.

*Data Collection and Preparation*

Any type of textual data can be used for TM. Data can be collected from online sources, such as company or organisation websites (Maier et al., 2018), customer reviews on products or services (Schmiedel et al., 2019), social media postings (Baumer et al., 2017), or offline sources, such as historical documents or archives (Croidieu and Kim, 2018), open-ended survey responses (Cintron and Montrosse-Moorhead, 2021), and existing research data collected for other purposes (Nelson, 2020). In line with the aim and objectives of this study, we have compiled our data from the PMI official website. The PMI provides reliable and valuable practical and academic publications on project management. Before starting the research, we confirmed that the PMI applies a peer-review process for all articles and an editorial check process for any other publications. Therefore, we chose the PMI website, for this study.

The PMI website consists of more than 15,000 textual documents, with publications categorised into around 45 knowledge areas and 22 industries. Since the primary concern of this study is "collaboration", we searched the term on the PMI search on-site search and filtered the case studies and whitepaper. Although the research's concern is high-tech industries such as aerospace or telecommunication, we did not exclude any industries. However, we included relevant areas of Complexity, Stakeholder Engagement, Strategy, Innovation, Governance, Portfolio Management, and Program Management. The search resulted in 33 case studies and 27 white papers. After removing 6 duplications and 1 literature review article, we considered 51 documents for analysis. These documents consist of sections such as abstract, introduction, literature review, methodology, findings and discussions, and conclusions. Since we have interested in data, results, and findings based on a real-life experience, we included only relevant sections. The final corpus consists of more than 400 pages and 250,000 words.

After our corpus is ready, we can start to perform the steps of the methodology. Figure 1 illustrates the steps of the proposed methodology. However, text analysis requires pre-processing to clean up the corpus, such as removing stop words, prefixes and suffixes, to make it ready for the algorithm to process (Nelson, 2020; Schmiedel et al., 2019). For this example, we used the KoNstanz Information MinEr (KNIME) Analytics Platform. With its build-in text processing package, KNIME allowed us to apply pre-processing and TM together (Villarroel Ordenes and Silipo, 2021). We developed our TM program on the platform and ran the algorithm to produce the raw codes. The proposed methodology is explained in more detail in the following.

*Figure 1 Steps of Proposed Methodology*

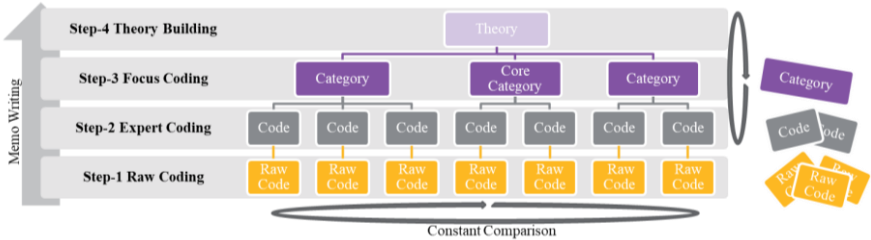

*Step 1 Raw Coding*

The first step is Raw Coding. A Raw Code is a TM topic. To build those raw codes, we suggest using the Latent Dirichlet Allocation (LDA) algorithm, which has proven to be a robust TM algorithm (Blei et al., 2003; Dimaggio et al., 2013; Lancichinetti et al., 2015; Maier et al., 2018).

To run LDA, we set the hyperparameters, such as the number of words in topics, and alpha and beta parameters, as suggested in the literature. Besides, the TM produced the same results for a given document cluster by keeping the parameters the same.

When the researcher gets their topics as raw codes, they need to make an initial analysis of them to prepare the raw codes for the next step of the methodology. Based on the nature of the data, some outlier codes would occur. Hence, the researcher would like to start with a higher number of topics to analyse but then continue with a smaller number (Croidieu and Kim, 2018). Thus, we are suggesting removing some outlier codes. For instance, although we were interested in high-tech industries, we did not remove any of the sectors during the search. We believe that each industry would include high-tech components. For example, "megaprojects" under the construction industry involve high-tech solutions. Although these documents include pure construction topics such as re-construction after an incident, they also discuss project management and collaboration related topics. Nevertheless, these construction discussing topics are not necessarily an error, but they need to be removed from the raw codes to move to the next stage of expert coding.

*Step 2 Expert Coding*
TM eases the text exploration process by reducing the complexity level of the text into a simpler or more interpretable set of words. However, looking at those words and making sense of it is time-consuming work that requires a special set of knowledge and capabilities a non-knowledgeable person would not possibly have. The Grounded Theorist researcher must have deep context knowledge of the area and content knowledge of the data. Hence, experts can make this process easy for the researcher. Experts can name topics (labelling) and rate the relevancy of a topic with their own experience (Kaplan and Vakili, 2015). These labels work as a title or a brief description for a raw code. Although experts would have sound contextual knowledge, they do not necessarily know about the research's scope. Therefore, before sharing raw codes, it is necessary to introduce the research to the experts and provide information about its background and objective. At the end of this process, raw codes turn into codes with a rate, and the researcher would like to remove low rated non-relevant codes.

*Step 3 Focus Coding*
In this step, the researcher engages with codes (and data if necessary) to seek and identify themes, patterns, and similarities within the codes. However, this is not a straightforward pattern recognition task for category building. Therefore, the researcher needs to look at the background meaning of the codes. Moreover, the labels would make the topics more understandable, so the researcher can develop categories. While developing categories, the researcher focuses on the most significant and/or frequent codes. The aim and objectives of the research, the knowledge gap identified with a literature review process, the background knowledge, the position, and the previous experience of the researcher become more visible in this step to develop these categories. Therefore, the researcher should involve the analytic tool of GT in this step. They engage in constant comparison to look at the similarities and differences (Charmaz, 2006). Eventually, with constant comparison, categories will start to emerge with links between codes. Before moving to the next stage, the researcher can remove the least frequent codes formed categories as suggested in CGT.

*Step 4 Theory Building*
The final step is theory building. This step is pretty much like the theoretical sampling of CGT. As Charmaz (2015) suggested, the goal of the researcher is to develop features of emerging categories to reach theoretical categories. She stated that it's not collecting and analysing a new set of data. Instead, she emphasized that it is theoretical sorting and integration to build meaningful links with constant comparison to frame the theory. Therefore, the researcher continues to apply constant comparison to get the idea or meaning behind the codes to reach analytical abstraction.

In this step, the researcher compares, contrasts and combines core categories to find links to build themes and new knowledge. However, identifying conceptual relations between categories would be a daunting job to do and requires critical thinking. The reason is, during this step, the researcher can make numbers of backwards and forwards, observations and reasoning between the steps, search for similarities and differences, and conceptual connections between the categories. To do so, the researcher is not just looking at patterns but building a link between pertinent categories. They constantly interact with codes, categories, and even data itself if needed during these steps. For example, since labels have a crucial role, the researcher would like to reach experts to understand the rationale of the label. Although this methodology perfectly serves the big data approach, the researcher would like to go deep inside the data during the third and fourth steps to enhance understanding. The LDA algorithm provides a document-topic matrix that allows seeing which document is represented by which topics, so one can read the document to see the whole picture for that particular code. This abstraction process should result in a real-life problem concerning middle-range theory (Easterby-Smith et al., 2018; Morse et al., 2021).

**Finding and Results**

The previous section introduced the steps of the proposed methodology. This section illustrates the initial results of each step.

We started with generating raw topics. We used hyperparameter settings for LDA as suggested in the literature (Griffiths and Steyvers, 2004). In particular, we set alpha to 0.5, beta to 0.02, and the number of words in a topic to 10. With regards to the number of topics, there is no consensus in the literature on a robust setting (Dimaggio et al., 2013). Unfortunately, the literature has not agreed on a solution to determine the right number of topics. While some scholars use error measuring-based algorithms to decide the optimum number of topics, most social science scholars either decide on a number (Croidieu and Kim, 2018) or generate several alternative numbers of topics (Baumer et al., 2017) and use the most relevant one for the analysis. For social sciences, the second approach for setting the number of topics is the most popular one (Dimaggio, 2015; Maier et al., 2018; Nelson, 2020). Since the proposed methodology uses the big data approach, this choice of the number of topics or words has not significantly changed the result. For example, when we changed the number of words from 10 to 12, the TM added 2 new relevant words to the existing ones.

In this example, we choose three alternatives for the number of topics (40, 50, and 60 topics). Then we compared topic groups and sought the similarity between 40-50 topics, 40-60 topics and 50-60 topics. We saw that the similarity of 40-50 and 40-60 are higher than the similarity of 50-60. While 85% and 87.5% of 40 topics were covered in 50 topics and 60 topics, respectively, 60 topics compromised 78% of 50 topics. Consequently, we decided to go with 40 topics. In line with the scope of the research, we cleaned the cluster of topics from the outlier code before sharing the codes with experts. We removed 6 topics from the cluster. **Error! Reference source not found.** illustrates the part of results for 40 topics.

*Table 1 Part of Results for 40 Topics*

| Topic Number | Words |
|---|---|
| topic_0 | team culture cultural manager program behavior rule quotient influence intelligence |
| topic_1 | management practice collaboration innovation project collaborative dynamic people manage agree |
| topic_2 | client delivery sponsor development megaproject supply chain organisation partner design |
| topic_3 | portfolio business strategic value organisation resource system information phase alignment |
| topic_4 | product company development budget technology approach market respondent review gate |
| ... | |

| ... | |
|---|---|
| topic_39 | team virtual difference virtually highly management development consider impact help |

TM topic generation relies on the nature of linguistics because words used together are related to a topic discussed in the content. Although these raw codes would tell the content of the data (project management and collaboration), they are not telling the meaning behind the words. It requires content and context knowledge. With this amount of data, one cannot get fully familiarized with the documents' content. However, experts can cover this requirement with contextual knowledge. We invited two project management experts for this task. The first expert has both academic and industrial project management experience of more than 20 years, and the second one has 10 years of industrial experience. Then we introduced the research to the experts and provided information about the research's background and objective. After that, we shared raw codes to benefit from their domain knowledge. We expected raw codes mean something to them so they could label the raw codes and rate their relevancy from 1 to 5. In the meantime, we also labelled the raw codes and wrote memos about our reasoning during the labelling process to understand the sense-making process. In the end, we aggregated all labels into one label (Table 2 for sample labels). The entire process unlocked the hidden meaning of the raw codes and allowed us to develop a further understanding of codes. We calculated the average of the relevancy rating and removed 4 codes where the rating average was less than 2. This process left us with 30 codes for the Focus Coding step.

*Table 2 from Raw Code to Aggregated Categories*

| Topic Number | Words | Label | Categories |
|---|---|---|---|
| topic_2 | client delivery sponsor development megaprojects supply chain organisation partner design | Stakeholder Collaboration for Innovation Project | Innovation and Stakeholder Engagement |
| topic_32 | bridge value stakeholder local business owner citizen impact sublime community | Stakeholder Collaboration for Value | |
| topic_0 | team culture cultural manager program behavior rule quotient influence intelligence | Management Influence on Organisational Adaptation | Leadership Involvement for Collaboration |
| topic_33 | project autonomy center manager unit stakeholder organisation network environment complex | complex Projects Governance, Autonomy and Collaboration | |

Then we started the Focus Coding step. We checked each code and visited documents when it was needed to understand the meaning of a code. We benefitted from words in topics, labels, memos, and documents. We applied the constant comparison tool of CGT to seek similarities and differences. After the discussion sessions, our categories started to emerge. Then we focused on the relevant categories that are significant to the research context and neglected non-relevant ones. During this process, we have developed 10 categories. We found that 4 of these categories were represented by a code. As suggested, we removed categories if they consist of one code as the least frequent categories. Additionally, we realized that one code could serve distinct categories during this process. To the best of our knowledge, this has not been discussed in any studies before. For example, Topic_20 can serve three distinct categories. There were four codes like that, so we included those codes under the relevant categories.

In our example, we identified two generic categories and called them "Project Management" and "Collaborative Innovation (CI) during a Project". Both categories were expected because of the analysed data nature. Nevertheless, they revealed that collaboration occurs at every project stage in every direction simultaneously. Moreover, we found four core categories for CI projects. These core categories are "Human Capital of CI", "Innovation and Stakeholder Involvement", "Leadership Involvement for Collaboration", and "Structural capital of CI". Then we aggregated those core categories into two themes: Capabilities for CI and Governance of CI. Besides, initial discoveries revealed the promising nature of the methodology. **Error!**

**Reference source not found.** illustrates the full list of topics and assigned categories with a higher-level construct.

*Table 3 Topics and Categories assigned to Aggregated Dimensions*

| Topic Number | Categories | Aggregate Dimensions |
|---|---|---|
| topic_0, topic_1, topic_20, topic_18, topic_21, topic_33, topic_36 | Leadership Involvement for Collaboration | Governance of CI |
| topic_15, topic_30, topic_31, topic_37, topic_39 | CI During a Project | Generic Category |
| topic_2, topic_13, topic_14, topic_17, topic_29, topic_30, topic_32 | Innovation and Stakeholder Involvement | Governance of CI |
| topic_20, topic_24, topic_25, topic_28, topic_34, topic_35 | Project Management | Generic Category |
| topic_3, topic_12 | Human Capital of CI | Capabilities for CI |
| topic_3, topic_4, topic_7, topic_8, topic_10, topic_17, topic_20, topic_27 | Structural capital of CI | Capabilities for CI |

**Conclusion**

This paper illustrated how an unsupervised machine learning textual analysis can be merged with a GT approach under a robust methodology. It combined the LDA algorithm with CGT in a novel way. We illustrated the methodology by applying it to collaboration concerned case studies on the PMI website and identified two theories for a complex innovation project, Governance of CI and Capabilities for CI. The governance of CI depends on stakeholders' engagement and leadership involvement in collaboration. Capabilities for CI rely on an organisation's tangible and intangible resources. Perhaps, these two themes can be merged into a higher-level theme and be called a CI Capability.

One other important aspect of research is quality. This paper utilised secondary data as PMI documents (Schmiedel et al., 2019) and primary data as expert-generated labels (Rinke et al., 2021). For this reason, both parts need to be discussed separately. For the secondary data, previous studies revealed that the TM algorithm fulfils the reliability and validity measures (Lancichinetti et al., 2015). For the primary, Kelle and Erzberger (2004) asserted that discussing the same phenomenon with diverse backgrounds and the expertise of experts provides different perspectives to the study, which fulfil the triangulation requirement for research. For example, Köhler et al. (2022) invited and conducted an interview with academics from the field who has varied backgrounds in terms of experience and expertise in the field to triangulate their findings. Therefore, we claim that expert labelling increases the study's credibility so the trustworthiness.

There are implications of our research. Management researchers would benefit from this novel methodology for exploratory research. This methodology allows them to analyse a vast amount of textual data in a short time. TM decreases a researcher's bias during the initial coding stage and identifies the hiding topics when a researcher does their own coding. Furthermore, expert involvement increases the credibility of the research. Additionally, it makes a novice researcher able to conduct a GT. Unlike the traditional GT approaches, where an experienced researcher with the contextual knowledge needed to conduct research, this methodology decreases this requirement.

There are, of course, aspects of the methodology that can be improved further. First, as Charmaz (2006) suggested, we have bravely faced our biases and conducted this research against our biases. It would be interesting to see another research team analyse the same data set to develop their theories. Secondly, we have collected our data from a trusted international organisation's website. However, as suggested previously, any textual data can be analysed with this methodology. Further research can collect and analyse data from relevant companies' publicly available websites. This time the sampling and data reliability issues would arise. That would be interesting to see how researchers solve those issues.